\documentstyle [12pt]{article}
\begin{document}
\begin{titlepage}
\rightline{March, 1993}
\vspace{0.3cm}
\rightline{KHTP-93-02}
\rightline{SNUCTP-93-17}
\vspace{0.3cm}
\centerline{\LARGE Classical $W_3^{(2)}$ algebra and its Miura map }
\vspace{1.5cm}
\centerline{\Large B.K.~Chung, K.G.~Joo, and Soonkeon~Nam\footnote{
E-mail address: nam@nms.kyunghee.ac.kr}}
\vspace{0.5cm}
\centerline {Research Institute for Basic Sciences}
\centerline {and Department of Physics,}
\centerline {~Kyung Hee University, ~Seoul, 130-701, ~Korea}
\vspace{4cm}
\centerline{\Large Abstract}
We verify that the fractional KdV equation is a bi-hamiltonian
system using the zero curvature equation in $SL(3)$ matrix valued Lax pair
representation, and explicitly find the closed form for the hamiltonian
operators of the system. The second hamiltonian operator is the classical
version of the $W_3^{(2)}$ algebra.
We also construct systematically the Miura map of $W_3^{(2)}$ algebra
using a gauge transformation of the $SL(3)$ matrix valued Lax operator
in a particular gauge, and construct the modified fractional KdV equation
as hamiltonian system.
\end{titlepage}

\def\be{\begin{equation}}
\def\ee{\end{equation}}
\def\bea{\begin{eqnarray}}
\def\eea{\end{eqnarray}}
\renewcommand{\arraystretch}{1.5}
\def\ba{\begin{array}}
\def\ea{\end{array}}
\def\bce{\begin{center}}
\def\ece{\end{center}}

\def\nn{\noindent}

\def\rhu{\frac{\delta H}{\delta u}}
\def\rhv{\frac{\delta H}{\delta v}}
\def\rhq{\frac{\delta H}{\delta q}}
\def\rhj{\frac{\delta H}{\delta j}}

\def\pbv{\partial_x^5}
\def\pbth{\partial_x^3}
\def\pbs{\partial_x^2}
\def\pbt{\partial_t}
\def\pbx{\partial_x}
\def\pbi{\partial_x^{-1}}

\def\dj{{\cal D}[j]}
\def\djj{{\cal D}[j_1,j_2]}
\def\dsh{{\cal D}[\psi,\phi]}
\def\du{{\cal D}[u]}
\def\dq{{\cal D}[q]}
\def\dfu{{\cal D}^{(1)}(x)}
\def\dsu{{\cal D}^{(2)}[u]}
\def\dfuv{{\cal D}^{(1)}[u,v]}
\def\dsuv{{\cal D}^{(2)}[u,v]}
\def\dshggt{{\cal D}^{(2)}[H,G^+,G^-,T]}
\def\dfhggt{{\cal D}^{(1)}[H,G^+,G^-,T]}
\def\df{{\cal D}^{(1)}}
\def\ds{{\cal D}^{(2)}}
\def\djjjj{{\cal D}[j_1,j_2,j_3,j_4]}

\def\fduj{\left[\frac{du}{dj}\right]}
\def\fduvjj{\left[\frac{d(u,v)}{d(j_1,j_2)}\right]}
\def\fdju{\left[\frac{dj}{du}\right]}
\def\fdjjuv{\left[\frac{d(j_1,j_2)}{d(u,v)}\right]}
\def\fdjjsh{\left[\frac{d(j_1,j_2)}{d(\psi,\phi)}\right]}
\def\fdshjj{\left[\frac{d(\psi,\phi)}{d(j_1,j_2)}\right]}
\def\fdqj{\left[\frac{dq}{dj}\right]}
\def\fdqu{\left[\frac{dq}{du}\right]}
\def\fduq{\left[\frac{du}{dq}\right]}
\def\fdhgtj{\left[\frac{d(H,G^+,G^-,T)}{d(j_1,j_2,j_3,j_4)}\right]}

\def\ho{hamiltonian operator}

In the study of integrable models\cite{FTDZ} as bi-hamiltonian systems, one
of the interesting thing is the connection between the integrable nonlinear
differential equations in two dimensions and the two dimensional conformal
field theories. The second hamiltonian structure of the $SL(N)$ extended KdV
equation corresponds to the classical form of the $W_N$ conformal
algebra\cite{Ge,Ma}.
The corresponding integrable hierarchy of nonlinear evolution equations is
characterized by a scalar Lax operator of order $N$, involving fields of
integer
conformal dimensions ranging from 2 to $N$\cite{DS,KW}. By the method of
hamiltonian reduction, the $W_N$ algebra is derived from an $SL(N)$ matrix
in a special gauge, in which all the elements of the strictly upper
triangular part are set equal to zero except those on the first diagonal,
where all elements are set equal to one\cite{Bo}. The classical $W_3$ algebra
of Zamolodchikov\cite{Za}, involving fields of spin-3 local current, can be
obtained
in the $SL(3)$ case and the corresponding integrable nonlinear differential
equation is the Boussinesq(Bsq) equation.
\newline\indent
 In studying two dimensional quantum gravity, Polyakov\cite{Pol} has noted
that one can choose a different gauge in the $SL(3)$ case, in
which the first diagonal elements of the strictly upper triangular part are
set equal to zero and the second diagonal element is set equal to one. Here,
in the so called $W_3^{(2)}$ algebra, two bosonic fields of spin-3/2
and a spin-1 field appear, in addition to the energy momentum tensor. This
suggests that there is a whole new class
of $W_N$ algebras denoted as $W_N^{(l)}$, obtained by setting the
elements of the $l$-th upper diagonal of an $SL(N)$ matrix to one and setting
the other elements in the upper triangular part to zero.
\newline\indent
As the $W_N$ algebra are directly related to the second hamiltonian structure
of $SL(N)$ extended KdV equation, it is a natural that this will
also be the case for the $W_N^{(l)}$ algebras\cite{BD,MO}.
It was shown in Ref.\cite{BD} that the new non-linear
evolution equations, called the fractional KdV equation, can be constructed by
using dimensional reduction of
self-dual Yang-Mills equations with two killing symmetries, and making
suitable ans\"{a}tz for the matrix components of the gauge connections.
These equation can be written
as hamiltonian systems with poisson brakets given by the classical $W_3^{(2)}$
algebra. These equations were found to be different from the Bsq equation, for
which the second hamiltonian structure is the standard $W_3$ algebra.
Such an equation can be also obtained from the Bsq equation by interchanging
the role of independent variables $x$ and $t$\cite{MO}.
Furthemore, the free field representation, i.e. the generalized Miura map, for
$W_N$ was
obtained either from a gauge tranformation\cite{Bo}, or a
factorization of the scalar Lax operator\cite{DS,KW}.
Introducing a two
component scalar field and two bosonic fields with spin-1/2, the free field
represention for $W_3^{(2)}$ algebra was also constructed \cite{BD,MO,Be}.
\newline\indent
In this paper we will verify the fractional KdV equation as a bi-hamiltonian
system using the zero curvature condition with $SL(3)$ matrix valued Lax pair
representation and explicitly find the closed form for the hamiltonian
operators of the system. The second hamiltonian operator is the classical
version of the $W_3^{(2)}$ algebra. From this, one can formulate the
non-linear
equations related to $W_N^{(l)}$ conformal algebras for
all the integer $N$ and $l$ ($1 \leq l \leq N-1$).
We will also construct systematically the Miura map of $W_3^{(2)}$ algebra
using a gauge transformation of the $SL(3)$ matrix valued Lax operator and
particular gauge choice, and verify the modified fractional KdV equation as
hamiltonian system. This method can be
extended to the general $W_N^{(l)}$ case.
\newline\indent
First of all, we will consider the well-known hamiltonian operators
and Miura map of the Bsq system in the zero curvature formulation. This will
serve to fix the notations we will be using in the case of the
fractional KdV system. We parametrize two gauge potentials of the $SL(3)$
matrix valued Lax pair operators as
\be \label{1}
U=\left(\begin{array}{ccc} 0 & 1 & 0 \\ 0 & 0 & 1 \\ u_1+\lambda & u_2 &
0\end{array}\right), \quad
V=\left( \begin{array}{ccc} P_1(x,t) & P_2(x,t) & P_3(x,t) \\ Q_1(x,t) &
Q_2(x,t) & Q_3(x,t) \\ S_1(x,t) & S_2(x,t) & S_3(x,t)\end{array}\right), \quad
Tr V=0,
\ee
where $\lambda$ is a constant parameter, and the
corresponding Lax operator is $L=\pbx-U$. Using
the zero curvature condition $[\pbx-U, \pbt-V]=0$, we obtain the
dynamical equations of $u_1$, $u_2$ and the constraint equations\cite{DHR}.
These
equations are expressed only in terms of the variables of upper triangular
part of $V$, i.e.
$P_2$, $P_3$ and $Q_3$.
Redefining  $u=u_2, \quad v=u_{2x}-2u_1$, we recover the standard
expression\cite{BD,MO}. Using one of the constraint equations,
$P_{3x}+P_2-Q_3=0$, and setting $\tilde{Q_3}=Q_3-\frac{1}{2}P_{3x}$,
$\tilde{P_3}=-\frac{1}{2}P_3$, we can write the hamiltonian equations of $u$,
$v$ variables in a matrix form:
\be \label{2}
\left(\begin{array}{c} u \\ v
\end{array}\right)_{\displaystyle t}=\left(\dsuv-6\lambda
\dfuv\right)\left( \begin{array}{c} \tilde{Q_3} \\ \tilde{P_3}
\end{array}\right),
 \ee
where the first hamiltonian operator, $\dfuv$, and
the second hamiltonian operator, $\dsuv$, are
\bea \label{3}
\dfuv\!\!\!\!&=&\!\!\!\!\left( \begin{array}{cc} 0 & \partial_x \\ \partial_x &
 0
\end{array}\right) \, , \nonumber \\
\dsuv\!\!\!\!&=&\!\!\!\!
\left( \begin{array}{cc} -2\pbth+2u\pbx + u_x & 3v\pbx +2v_x \\ 3v\pbx +
v_x  & {{\displaystyle 2{\pbv}/3-10u{\pbth}/3-5u_x \pbs} \atop
{\displaystyle +(8u^2/3 -3u_{xx})\pbx -2u_{xxx}/3 + 8uu_{x}/3}}
\end{array}\right).
\eea
When we put the upper triangular part of $V$ as the dual of the upper
triangular part of $U$, i.e. $P_2=Q_3=0$, $P_3=1$ ($\tilde{Q_3}=0$,
$\tilde{P_3}=-1/2$), we obtain the Bsq equation,
$$
u_t=-v_x, \quad v_t=\frac{1}{3}u_{xxx} -\frac{4}{3}uu_x.
$$
The other elements of $V$ are obtained in terms of $u$, $v$
variables by the constraint equation.
The hamiltonian operators of eq.(\ref{3}) define the poisson structure of
Bsq equation.
The poisson brackets represented by the second hamiltonian operator,
\be \label{4}
\left( \begin{array}{cc} \{u(x), u(y)\} & \{u(x), v(y)\}  \\ \{v(x), u(y)\} &
\{v(x), v(y)\} \end{array}\right)=\dsuv\delta(x-y),
\ee
is the classical version of $W_3$ algebra\cite{Ma}.
\newline\indent
To look for the Miura map of the Bsq equation, that is the free field
representation of the above $W_3$ algebra, we perform the gauge
transformation $L\longrightarrow\Phi^{-1}L \Phi$ for the Lax operator $L$ with
$\Phi(x,t)$ taking values in the strictly lower triangular matrices of $SL(3)$
with the diagonal elements set equal to one. Setting the transformed Lax
operator as
$\tilde{L}=\pbx-\tilde{U}$, where $\tilde{U}=\Phi^{-1}U \Phi-\Phi^{-1}\Phi_x$,
and choosing a gauge as
\be
\tilde{U}=
\left( \begin{array}{ccc} j_1 & 1 & 0  \\ 0 & -(j_1 + j_2) & 1 \\ \lambda
& 0 & j_2 \end{array}\right) ,
\ee
we obtain the Miura map of $u$ and $v$ with respect to $j_1$ amd $j_2$,
\bea \label{5}
u\!\!\!\!&=&\!\!\!\!(j_{1}-j_{2})_x + j_{1}^{2} + j_{1}j_{2} +
j_{2}^{2}, \nonumber\\
v\!\!\!\!&=&\!\!\!\!-(j_1 + j_2)_{xx} -(2j_1 - j_2)j_{1x} + (2j_2 - j_1)j_{2x}
+ 2(j_1
+ j_2)j_{1}j_{2}.
\eea
The modified Bsq equation(mBsq) for $j_1$, $j_2$ and its hamiltonian structure
can be exhibited by the Fr\'{e}chet derivative of the transformation of
eq.(\ref{5}).
In general, given a transformation $u=F[j, j_{x}, \cdots ]$,
the Fr\'{e}chet derivative, ${\displaystyle \fduj}$, is
the differential operater that implies $u_t={\displaystyle \fduj} j_t$ and for
any functional $H[u]$, we have
${\displaystyle \rhj}={\displaystyle {\fduj}^{\ast} \rhu}$, where
${\displaystyle {\fduj}^{\ast}}$
is the formal adjoint of ${\displaystyle \fduj}$.
The Fr\'{e}chet derivative and its formal adjoint with respect to the
transformation eq.(\ref{5}) can be written in a matrix form as
\bea \label{7}
\fduvjj\!\!\!\!&=&\!\!\!\!\left( \begin{array}{cc} \pbx+2j_1+j_2 &
-\pbx+2j_2+j_1 \\
{{\displaystyle -\pbs+(j_2-2j_1)\pbx-2j_{1x}}\atop
{\displaystyle -j_{2x}+2j_{2}^{2}+4j_{1}j_{2}}}
& {{\displaystyle -\pbs+(2j_2-j_1)\pbx+j_{1x}}\atop
{\displaystyle+2j_{2x}+4j_{1}j_{2}+2j_{1}^{2}}}
\end{array}\right),  \nonumber \\
{\fduvjj}^{\dagger}\!\!\!\!&=&\!\!\!\!\left( \begin{array}{cc} -\pbx+2j_1+j_2 &
{-\pbs-(j_2-2j_1)\pbx -j_{2x}+2j_{2}^{2}+4j_{1}j_{2}} \\ \pbx+2j_2+j_1 &
{-\pbs-(2j_2-j_1)\pbx+2j_{1x}+4j_{1}j_{2}+2j_{1}^{2}}
\end{array}\right).
\eea
The hamiltonian operator of mBsq equation,$\djj$, and the second
hamiltonian operator of Bsq equation, $\dsuv$, are related by,
through the Fr\'{e}chet derivative and its formal adjoint,
\be \label{6}
\dsuv=\fduvjj \djj {\fduvjj}^{\dagger},
\ee
and it is straightforward to show that $\djj$ takes the form
\be \label{8}
\djj =\frac{1}{3}\left( \begin{array}{cc} 2\pbx & -\pbx \\ -\pbx & 2\pbx
\end{array}\right).
\ee
(This operator can be also found
introducing the inverse Fr\'{e}chet derivative\cite{Joo}.) This hamiltonian
operator defines the poisson structure of mBsq equation as eq.(\ref{4}) in
the case of Bsq equation. The mBsq equation for $j_1$,$j_2$ is
\be \label{9}
{\left(\begin{array}{c} j_1 \\ j_2 \end{array}\right)}_{\displaystyle t}=
 \djj {\fduvjj}^{\dagger}
{\left(\begin{array}{cc} {{\displaystyle\frac{\delta}{\delta u}}} \\
{\displaystyle{\frac{\delta}{\delta v}}} \end{array}\right)} H= \frac{1}{3}
{\left( \begin{array}{c} (j_{1x}+2j_{2x}+j_{1}^{2}-2j_{2}^{2}-2j_1 j_2 )_x \\
(-2j_{1x}-j_{2x}-2j_{1}^{2}+j_{2}^{2}-2j_1j_2)_x \end{array}\right)},
\ee
where we have used ${\displaystyle \rhu}=\tilde{Q_3}=0$, ${\displaystyle\rhv}=
\tilde{P_3} = -\frac{1}{2}$ in eq.(\ref{2}).
\newline\indent
Through above methods, we can new find the fractional KdV
equation as a bi-ha\-mil\-ton\-ian system and show that the second hamiltonian
operator is the classical version of $W_3^{(2)}$ algebra, and we also find
explicitly the Miura map.
Firstly, we parametrize two gauge potentials of $SL(3)$ matrix valued Lax pair
operators as follows:
\be \label{10}
U=\left( \begin{array}{ccc} \frac{1}{2}H(x,t) & 0 & 1 \\ G^{+}(x,t) &
-H(x,t) & 0 \\ D(x,t)+\lambda & G^{-}(x,t) &
\frac{1}{2}H(x,t)\end{array}\right), \;\;
V=\left( \begin{array}{ccc} P_1(x,t) & P_2(x,t) & P_3(x,t) \\ Q_1(x,t) &
Q_2(x,t) & Q_3(x,t) \\ S_1(x,t) & S_2(x,t) & S_3(x,t)\end{array}\right),
\ee
with $Tr V=0$, where $\lambda$ is a constant parameter.
The zero curvature condition, $[\pbt-U, \pbx-V]=0$, gives five constraint
equations for $S_1$, $S_2$, $S_3$, $Q_1$ and $Q_2$, and four dynamical
equations for $H$, $G^+$, $G^-$ and $D$.
If we redefine $D+\frac{3}{4}H^2=T$,
the four dynamical equations for the variables	$H$, $G^+$, $G^-$ and
$T$ are expressed in terms of $P_1$, $P_2$, $P_3$ and $Q_3$ after
using the constraint equation as
\bea \label{12}
H_t\!\!\!\!&=&\!\!\!\!2P_{1x}+G^+ P_2-G^- Q_3+P_{3xx}\ ,\nonumber \\
G^+_t\!\!\!\!&=&\!\!\!\!-3G^+
P_1-Q_{3xx}-3HQ_{3x}+(\lambda+T-\frac{3}{2}H_x-3H^2)Q_3+G^+_x
P_3+\frac{3}{2}G^+ H P_3\ , \nonumber \\
G^-_t\!\!\!\!&=&\!\!\!\!3G^-
P_1+P_{2xx}-3HP_{2x}+(-\lambda-\frac{3}{2}H_x+3H^2-T)P_2+G^-_x P_3 \nonumber \\
& & \quad +3G^-P_{3x}-\frac{3}{2}HG^-P_3 ,  \\
T_t\!\!\!\!&=&\!\!\!\!3HP_{1x}+\frac{3}{2}G^+
P_{2x}+\frac{1}{2}G_x^+ P_2+\frac{1}{2}G_x^-
Q_3+\frac{3}{2}G^-Q_{3x}-\frac{1}{2}P_{3xxx} \nonumber \\
& & \quad +\frac{3}{2}HP_{3xx}+2(\lambda+T-\frac{3}{4}H^2)P_{3
x}+(T_x-\frac{3}{2}HH_x)P_3 \ .\nonumber
\eea
Setting $\tilde{P_1}=3P_1+\frac{3}{2}P_{3x}-\frac{3}{2}HP_3$, we can show that
above equtions express the hamiltonian equations for the variables $H$,
$G^{(\pm)}$ and $T$ in a closed form.
Splitting eq.(\ref{12}) into two parts, one that depends on the constant
parameter $\lambda$ and the rest, one can define two hamiltonian operators,
$\df$ and $\ds$.
That is, the eq.(\ref{12}) reads in a compact form as follows:
\be \label{13}
\Psi^{\dagger}_{t} =
\left(\dshggt + \lambda \dfhggt \right){\cal P}^{\dagger} \, ,
\ee
where $\Psi=(H, G^+, G^-, T)$,${\cal P}$=$(\tilde{P_1}, P_2, Q_3, P_3)$,
and the subscript $t$ denotes the time derivative. Here the matrix ${\cal P}$
means the functional variation for a hamiltonian,${\cal H}$, of the equation,
i.e. $\tilde{P_1}={\displaystyle \frac{\delta}{\delta H}}{\cal H}$,
$P_2={\displaystyle \frac{\delta}{\delta G^+}}{\cal H}$,
$Q_3={\displaystyle \frac{\delta}{\delta G^-}}{\cal H}$ and
$P_3={\displaystyle \frac{\delta}{\delta T}}{\cal H}$.
The first hamiltonian operator
$\df$ has non-vanishing elements ${\df}_{23}=1$, ${\df}_{32}=-1$,
and ${\df}_{44}=2\pbx$, and the second hamiltonian operator $\ds$ is
\be \label{14}
\ds=\left( \begin{array}{cccc} \frac{2}{3} \pbx & G^+ & -G^- & H_x+H \pbx
\\ -G^{+} & 0 &{{\displaystyle -\pbs-3H\pbx+T}\atop
{\displaystyle -3H_x/2-3H^2}} & \frac{3}{2}G^+\pbx+G_x^+ \\
G^{-} & {{\displaystyle \pbs -3H\pbx-T} \atop {\displaystyle -3H_x/2+3H^2}}
& 0 & \frac{3}{2}G^-\pbx+G^-_x	\\
H\pbx & \frac{3}{2}G^+\pbx+\frac{1}{2}G^+_x &
\frac{1}{2}G_x^-+\frac{3}{2}G^-\pbx & -\frac{1}{2}\pbth+2T\pbx+T_x
\end{array}\right) .
\ee
Putting ${\cal P}$=(0,1,1,0) in eq.(\ref{13}),
the other elements of $V$ are fixed by the constraint equations as
$P_1=Q_2=S_3=0$, $Q_1=S_2=-{\frac {3}{2}}H$ and
$S_1={\frac{1}{2}}(G^{+}+G^-)$,
and we can recover the fractional KdV equation of Ref.\cite{BD}, which we will
call the `first' hamiltonian equation, and see that
the hamiltonian of the equation is ${\cal H}=\int (G^{+}+G^{-})\,dx$.
In this particular case, the
upper triangular part of $V$ is a dual of the upper triangular part of $U$.
Returning to the general form of $\df$ and $\ds$,
the non-vanishing poisson brackets of the first hamiltonian structure
can be written as
$$
\{G^+(x), G^-(y)\}=\delta(x-y), \quad \{T(x), T(y)\}=2\pbx\delta(x-y),
$$
and for the second structure they are as follows:
\bea \label{15}
\{H(x), H(y)\}\!\!\!\!&=&\!\!\!\!\frac{2}{3}\delta(x-y),\quad
\{T(x), H(y)\}=H(x)\pbx\delta(x-y)  \ , \nonumber \\
\{H(x), G^{(\pm)}(y)\}\!\!\!\!&=&\!\!\!\!{\pm}G^{(\pm)}(x)\delta(x-y), \quad
\{G^{(\pm)}(x), G^{(\pm)}(y)\}=0 \ , \nonumber \\
\{T(x), G^{(\pm)}(y)\}\!\!\!\!&=&\!\!\!\!
(\frac{3}{2}G^{(\pm)}(x)+\frac{1}{2}G^{(\pm)}(x))\delta(x-y) , \\
\{T(x),
T(y)\}\!\!\!\!&=&\!\!\!\!(-\frac{1}{2}\pbth+2T(x)\pbx+T_x(x))\delta(x-y)
\ , \nonumber \\
\{G^+(x),G^-(y)\}\!\!\!\!&=&\!\!\!\!(-\pbs-3H(x)\pbx+T(x)-\frac{3}{2}H_x(x)
-3H^2(x))\delta(x-y) \ . \nonumber
\eea
These poisson brackets are the classical version of $W_3^{(2)}$ algebra
involving an abelian spin-1 current $H$, a spin-2 field $T$ (stress-energy
tensor)
and two bosonic spin-3/2 fields $G^{(\pm)}$, which were originally introduced
by
Polyakov\cite{Pol} and further investigated by Bershadsky\cite{Be}.
\newline\indent
It is important that find the Miura map of hamiltonian
equation(\ref{13}), i.e. the
free field representation, if we want to quantize the hamiltonian system.
We now look for the Miura map of fractional KdV equation systematically and
show that
it is the free field realization of the above second poisson brackets, also
find the modified fractional KdV equations as hamiltonian system. As in the
case
of $W_3$, we perform the gauge transformation for the Lax operator $L$ with
$\Phi(x,t)$ taking values in the strictly lower triangular matrices
of $SL(3)$ with ones on the diagonal. The gauge potential $\tilde{U}$ of the
transformed Lax operator $\tilde{L}$ is
$\tilde{U}=\Phi^{-1}U \Phi- \Phi^{-1}{\Phi}_x $, and when we choose a gauge as
\be \label{16}
\tilde{U}=
\left( \begin{array}{ccc} j_3 & j_1 & 1  \\
0 & -(j_3 + j_4) & j_2 \\
\lambda& 0 & j_4 \end{array}\right) \ ,
\ee
we obtain the Miura map;
\bea \label{17}
H\!\!\!\!&=&\!\!\!\!j_1j_2+j_3+j_4, \quad
T=j_3^2+j_4^2+j_3j_4-\frac{1}{2}j_1j_{2x}+\frac{1}{2}j_2j_{1x}+
\frac{1}{2}j_{3x}-\frac{1}{2}j_{4x}\ , \nonumber \\
G^+\!\!\!\!&=&\!\!\!\!-j_{2x}-2j_2j_3-j_1j_2^2-j_2j_4, \quad
G^-=j_{1x}-j_1^2j_2-2j_1j_4-j_1j_3 \ .
\eea
We can now see that $j_1, j_2$ are bosonic fields of spin half, and $j_3, j_4$
have spin 1.
Here we can clearly see the advantage of the redefinition of
$D+{\frac{3}{4}}H^{2}=T$. In terms of $j$'s $D$ has quartic terms, whereas
$T$ requires only those quadratic in $j$'s. The latter is suitable for
free field representation, which is easier to quantize.
In fact, to show that the above Miura map is the free field realization of
$W^{(2)}_{3}$ algebra, we must find the poisson structure of $j$'s.
\newline\indent
The hamiltonian operator of the modified fractional KdV
equation,$\djjjj$, which defines the poisson structure as before, and the
second hamiltonian operator of the fractional
KdV are related by, through the Fr\'{e}chet derivative and its formal adjoint,
\bea \label{18}
\dshggt=\fdhgtj \djjjj {\fdhgtj}^{\dagger} \ .
\eea
The Fr\'{e}chet derivative and its formal adjoint with respect to the
transformation eq.(\ref{17}) are explicity calculated as :
\bea \label{19}
\fdhgtj\!\!\!\!&=&\!\!\!\!
\left( \begin{array}{cccc} j_2 & j_1 & 1 & 1 \\
-j_2^2 & {{\displaystyle -\pbx-2j_3} \atop {\displaystyle -2j_1j_2-j_4}}
& -2j_2 & -j_2 \\
{{\displaystyle \pbx-2j_1j_2} \atop {\displaystyle -2j_4-j_3}}
& -j_1^2 & -j_1 & -2j_1 \\
-\frac{1}{2}j_{2x}+\frac{1}{2}j_2\pbx & -\frac{1}{2}\pbx+\frac{1}{2}j_{1x} &
{{\displaystyle \pbx /2} \atop {\displaystyle +2j_3+j_4}} &
{{\displaystyle -\pbx /2} \atop {\displaystyle +2j_4+j_3}}
\end{array}\right) \, , \nonumber \\
& & \nonumber \\
{\fdhgtj}^{\dagger}\!\!\!\!&=&\!\!\!\!
\left( \begin{array}{cccc} j_2 & -j_2^2
& {{\displaystyle  -\pbx-2j_1j_2} \atop {\displaystyle -2j_4-j_3}}
& j_{2x}-\frac{1}{2}j_2\pbx \\
j_1 & {{\displaystyle \pbx-2j_3} \atop {\displaystyle -2j_1j_2-j_4}} & j_1^2
& \frac{1}{2}j_1\pbx+j_{1x}\\
1 & -2j_2 & -j_1 & 2j_3+j_4-\frac{1}{2}\pbx  \\
1 & -j_2 & -2j_1 & 2j_4+j_3+\frac{1}{2}\pbx
\end{array}\right) \, .
\eea
Using eq.(\ref{18}) and the second hamiltonian operator of
eq.(\ref{14}), it is straightforward to show that the hamiltonian operator of
modified fractional KdV equation is
\bea \label{20}
\djjjj=
\left( \begin{array}{cccc} 0 & 1 & 0 & 0 \\ -1 & 0 & 0 & 0 \\
0 & 0 & \frac{2}{3}\pbx & -\frac{1}{3}\pbx
\\ 0 & 0 & -\frac{1}{3}\pbx & \frac{2}{3}\pbx
\end{array}\right)\, .
\eea
In particular we can see that the hamiltonian operator of mBsq equation
eq.(\ref{8}) is exactly involved in it of modified fractional KdV equation.
The above operator defines the poisson brakets of the modified
fractional KdV system, of which non-vanishing ones are
\begin{eqnarray*}
\{j_1(x), j_2(y)\}\!\!\!\!&=&\!\!\!\!\delta(x-y) \, , \quad
\{j_3(x), j_3(y)\}=\frac{2}{3}\pbx\delta(x-y) \, , \nonumber \\
\{j_3(x), j_4(y)\}\!\!\!\!&=&\!\!\!\!-\frac{1}{3}\pbx\delta(x-y) \, , \quad
\{j_4(x), j_4(y)\}=\frac{2}{3}\pbx\delta(x-y)\, .
\end{eqnarray*}
Setting $\phi_1=\frac{\sqrt{3}}{2}(j_3+j_4), \phi_2=\frac{1}{2}(j_3-j_4)$,
we can obtain
the poisson brackets in terms of $j_1$, $j_2$ and $\phi_1$, $\phi_2$ that
express the free field realization.
\newline\indent
Finally, we find the modified fractional KdV equation as hamiltonian
system using the hamiltonian operator eq.(\ref{20}) and the Fr\'{e}chet
derivative eq.(\ref{19}). The hamiltonian equation for $j$'s are
\be
J^{\dagger}_{t} =\djjjj {\fdhgtj}^{\dagger}{\Delta}^{\dagger} {\cal H} \, ,
\ee
where $J=(j_1,j_2,j_3,j_4)$, $\Delta=({\displaystyle \frac{\delta}{\delta H}},
{\displaystyle \frac{\delta}{\delta G^+}},
{\displaystyle \frac{\delta}{\delta G^-}},
{\displaystyle \frac{\delta}{\delta T}},)$ and ${\cal H}$ is a conserved
quantity of the fractional KdV equation, therefore $\Delta {\cal H}$ is the
very ${\cal P}$ in eq.(\ref{13}). When we take $\Delta {\cal H}$=${\cal P}$=
$(0,1,1,0)$ which give the fractional KdV equation of Ref.\cite{BD}, we can
obtain a modified fractional KdV equation following as
\bea \label{21}
j_{1t}\!\!\!\!&=&\!\!\!\!-j_1^2-2j_3-2j_1j_2-j_4 \, , \quad
j_{2t}=j_2^2+2j_1j_2+j_3+2j_4 \, , \nonumber \\
j_{3t}\!\!\!\!&=&\!\!\!\!-2j_{2x}, \quad  j_{4t}=-2j_{1x} \, .
\eea
Using the above-mentioned variables ${\phi}_1$, ${\phi}_2$ and resetting
$p_1=\frac{1}{\sqrt{2}}(j_1-j_2)$, $p_2=\frac{1}{\sqrt{2}}(j_1+j_2)$,
these equations agree with the equation (called\,  ${\widetilde {mBsq}}$)
given in Ref.\cite{MO} except the factors.
\newline\indent
As a final remark, we will discuss a systematic description of the
$W_N^{(l)}$ algebras and its Miura map for all $N$ and $l$ $(1\leq l\leq
N-1)$. For this we have to parametrize the gauge potential
defining the $SL(N)$ matrix valued Lax operator $U$, similar to
$U$ of eq.(\ref{1}) and
eq.(\ref{10}) for the case of $SL(3)$, where the elements of the
$l$-th upper diagonal part are set equal to one
and the other elements in the upper triangular
part to zero\cite{MO}. However this parametrization for $N>3$ have not
been explicitly constructed yet. The zero curvature condition expresses
dynamical
equations for the elements of $U$ and constraint equations for the elements
of $V$. Using the constraint
equations through an appropriate change of variables in $V$, the dynamical
equations will give hamiltonian equations, and the hamiltonian operator will
represent the classical $W_N^{(l)}$ algebra. As a particular case, when we
take the upper triangular part of $V$ same as the upper triangular part of
$U$, i.e. if we make use of the $x \leftrightarrow t$ duality, the
hamiltonian equation will
represent the `first' non-linear differential equation of corresponding
algebra.
The Miura map of this non-linear differential equation represents the free
field
realization of corresponding algebra. It is well known that the Miure map for
$W_N$ algebra ($l=1$ in $W_N^{(l)}$ case) is nothing but the relation
between two different choice of gauge slice\cite{DS,Bo}. In this paper we
also found the Miura map for $W_3^{(2)}$ case using an appropriate gauge
choice. Thus, we expect such a procedure will also gives Miura maps of general
$W_N^{(l)}$.
\newline\indent
We would like to thank Q-Han Park for discussions.
This work was supported in part by Ministry of Education, and by Korea Science
and Engineering Foundation.
%

\end{document}